\documentclass[%
 aip,
 jmp,%
 amsmath,amssymb,
 reprint,%
]{revtex4-2}

\usepackage{graphicx,color,mhchem}
\usepackage{dcolumn}
\usepackage{bm}

\begin{document}
\title[UV-enhanced SEM: towards orientation and electron work function imaging]
{UV-enhanced SEM: towards orientation and electron work function imaging}

\author{Maciej Kretkowski}%
\affiliation{Research Institute of Electronics, Shizuoka University, 3-5-1 Johoku, Chuo-ku, Hamamatsu 432-8011, Shizuoka, Japan}%

\author{Haoran Mu}%
\affiliation{Optical Sciences Centre, School of Science, Swinburne University of Technology, Hawthorn, Victoria 3122, Australia }%

\author{Hsin-Hui Huang}%
\affiliation{Optical Sciences Centre and ARC Training Centre in Surface Engineering for Advanced Materials (SEAM), School of Science, Swinburne University of Technology, Hawthorn, VIC 3122, Australia}%

\author{Krishna Prasad Khakurel}%
\affiliation{Extreme Light Infrastructure (ERIC), Za Radnici 835, 25241 Dolni Brezany, Czechia}%

\author{Lukita Sari Ikhsan}%
\affiliation{Graduate School of Science and Technology, Shizuoka University, 3-5-1 Johoku, Chuo-ku,Hamamatsu 432-8011, Shizuoka, Japan}%

\author{Yu Masuda}%
\affiliation{Faculty of Engineering, Shizuoka University, Hamamatsu, Shizuoka 432-8561, Japan}%

\author{Saulius Juodkazis}%
\affiliation{Optical Sciences Centre, School of Science, Swinburne University of Technology, Hawthorn, Victoria 3122, Australia }
\affiliation{World Research Hub Initiative (WRHI), Institute of Science Tokyo, School of Materials and Chemical Technology, 2-12-1, Ookayama, Meguro-ku, Tokyo 152-8550, Japan}
\affiliation{Laser Research Center, Physics Faculty, Vilnius University, Saul\.{e}tekio Ave. 10, 10223 Vilnius, Lithuania}

\author{Wataru Inami}%
\affiliation{Research Institute of Electronics, Shizuoka University, 3-5-1 Johoku, Chuo-ku, Hamamatsu 432-8011, Shizuoka, Japan}%

\author{Yoshimasa Kawata}%
\affiliation{Research Institute of Electronics, Shizuoka University, 3-5-1 Johoku, Chuo-ku, Hamamatsu 432-8011, Shizuoka, Japan}%

\date{\today}
            
\begin{abstract}
Deep-UV $\sim 250$~nm (4.96~eV) tilted \emph{in-situ} co-illumination of the sample under imaging by scanning electron microscope (SEM) is developed at a robust and practical instrument level. Precise mechanical control of the lateral position and tilt angle (within 6.5$^\circ$ from a 42$^\circ$ baseline) of the UV-C LED source is achieved using mechanisms external to the vacuum chamber. The incorporated linear polariser (for s-pol. mode illumination) with external polarisation plane adjustment allows for modulation and tuning of tangential $E^{(t)}$ and normal $E^{(n)}$  electric field components and their enhancement for controlled directional electron emission from the surface of the sample. Numerical modelling of E-field enhancement corroborates the expected enhancement in the production of secondary electrons. This modality of SEM imaging does not require metal coatings, preserving sample integrity for subsequent analysis. The feasibility of having linearly polarised incident UV-C light with azimuthal orientation control in $(s,p)$-plane is modeled and discussed. 
\end{abstract}

\maketitle


\section{\label{intro}Introduction}

As microelectronics is entering the 2-nm-mode of lithography, the scanning electron microscopy (SEM) needs higher resolution for imaging, which is usually achieved by higher accelerating voltage. This leads to increased generation of secondary electrons and possible structural damage of, e.g.,  minute  field transistors $\sim 10-30$~nm in the cross section. SEM requires a metal coating of at least 2-3~nm for imaging, which renders imaged micro-electronics items not functional. One possible solution is to illuminate the sample under SEM inspection with deep UV light 240-280~nm, which corresponds to the electron work function, $w_e$, of most materials. Lower energy quanta are required to free the electrons from shallow tarps and  surface states.  Another burgeoning field of optical quantum emitters formed in SiC, N-doped diamonds, GaN, etc. needs surface characterisation with SEM field emission $\sim 1$~nm resolution without metal coatings. For example, the surface of a wide bandgap $5.48$~eV (226~nm) diamond is strongly charging under SEM imaging, which can be mitigated by using deep-UV illumination. The electron work function for diamond is $w_e = 4-5$~eV and can be greatly reduced for the n-type with N-doping (N created $1.7$~eV electron trap below the conduction band edge), graphitised diamond has $w_e = 4.4-4.7$~eV. 

Although the idea of using deep-UV for mitigation of surface charging in electron and ion beam imaging was demonstrated on the conceptual level~\cite{13lpr1049,16aplp021301}, no actual instrumentation solution is available from major vendors of such instruments. As samples without metal coating can be SEM imaged using UV-C co-illumination~\cite{13lpr1049,16aplp021301}, the prospects for application are highly desirable. The prototype of UV-C illuminator addition to a SEM microscope was recently demonstrated~\cite{Maciek_UV_base} and showed up to $50\%$ enhancement in secondary electrons for Si imaging under 250~nm co-illumination which already enables electron generation via photo-effect.    

The field of SEM is approaching 100 years mark counting from 1931 when M.Knoll and E. Ruska invented the microscope. It was later adopted for imaging in atmospheric conditions, hence, ASEM~\cite{Danilatos,Suga}, where, e.g., cells can be imaged through a thin membrane in a hydrated state. Furthermore, combining atomic force microscopy (AFM) with SEM is allowing to achieve 3D rendering of SEM images~\cite{afmsem} with advantage of probing mechanical properties.
While the research in SEM modality is advancing steadily within biomedical and hybrid applications (like D-EXA~\cite{Maciek_EXA_turret,MK_Vacuum_Gripper}, Cryo-SEM~\cite{GUAITA2022102484} , 4D-STEM~\cite{LIU2026906} etc.), the UV-enhanced mode remains an emerging and largely unexplored frontier. This paper aims to conceptualize the potential of this technique, supported by a practical implementation using a highly configurable, custom-engineered UV-C illumination instrument (module).

The technical challenges in the practical application of the UV illumination inside SEM revolve around vacuum compatibility, in-situ electrical/mechanical control, geometrical constraints of the vacuum chamber, and the necessity of direct irradiation with minimal optical power losses. Strict requirements regarding material out-gassing, restricted space inside the observation chamber, mechanical proximity to detectors, etc. are good examples of limiting factors. Moreover, the multi-disciplinary design expertise and familiarity with SEM systems in conjunction with optics limits the availability of vendors, suppliers, and designers of auxiliary equipment. Consequently, the inherently high costs required to develop a custom deep UV illuminator for research purposes may be a deterring factor.

Here, the practical solution for a deep-UV co-illumination during SEM imaging~\cite{Maciek_UV_base} is detailed, revealing the technical implementation of control in lateral/axial positions and the tilt angle of the UV-C illumination. Augmentation of SEM capabilities under UV-C co-illumination is conceptualised by introducing a linear polariser after UV LED into the same side-port attachable instrument. Optical performance in terms of light field enhancement is numerically simulated for 3D nano-scale patterns composed of dielectric, semiconductor, and metal.   
\begin{figure}[b!]
\centering\includegraphics[width=\textwidth]{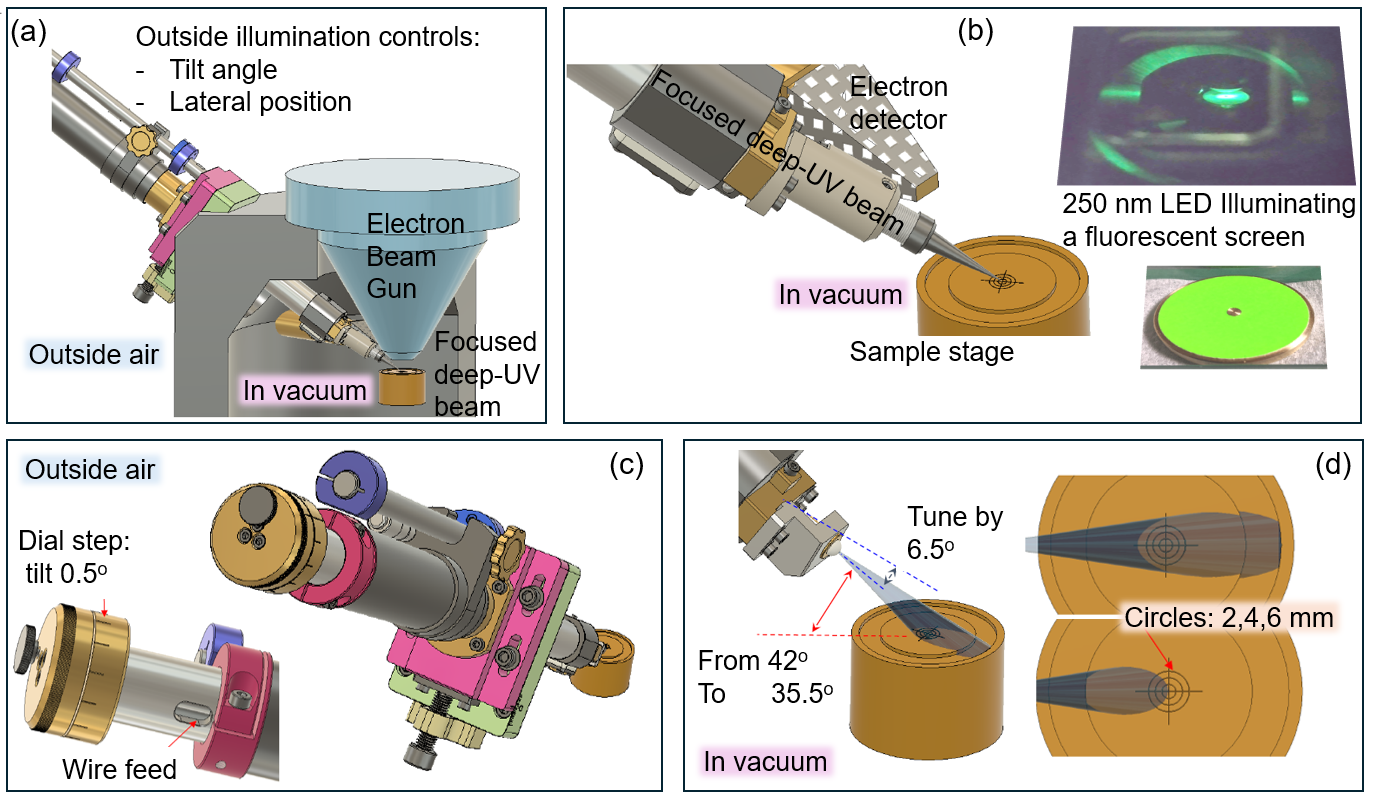}
\caption{\label{f-jig} Technical implementation of the deep-UV module combined with SEM (FE Jeol 7001). (a) The module interfaced with the SEM's sample area via side-port. (b) Closeup view of sample stage and actual optical image of the 250~nm LED illuminated sample holder. (c) The outside controls of lateral position and tilt of illumination detailed in (d). See details on  focusing of the UV light in the text and in Fig.~\ref{f-illum}. }
\end{figure}

\section{Design}

Figure~\ref{f-jig} shows the 3D CAD design of the deep-UV module for installation on the standard FE Jeol 7001 field emission SEM. The inset of (a) shows the principle of operation of the instrument and its alignment in relation to the Electron Beam Gun and the sample holder. The design is focused on intermittent use of the device by bringing it to the pre-set proximity of the measured sample when in use, and withdrawal of the end effector (illuminator) to its resting position during standard operation of the SEM. This practical solution allows for comparative studies and uninterrupted installation of other auxiliary modules such as CL mirrors, optical fibers, sensors, etc.~\cite{TROYON199815,Nedela}.
In our study, we focus on the direct illumination approach of the UV beam to the surface of the sample, thus avoiding the optical power losses typically present in typical optical fiber-based solutions~\cite{Raptis:16}. The aforementioned pre-set proximity position (fine-adjustable by means of a micrometer screw) allows for adjustment of the angle of illumination as well as the distance of the UV source from the sample. In the inset of (b), we show the focused illumination and the actual image of a fluorescent paper response to the focused beam.The challenges in designing such instrument are closely related with the vacuum requirements (in order of $5\times 10^{-5}$~Pa and limited space inside the SEM's vacuum chamber as dictated by the geometry of the microscope and various internal components (E-gun, sensors, stage control mechanism, etc.). Furthermore, the electrical contacts of the UV LED as well as the tilt and proximity controls have to be accessible from the outside of the vacuum chamber, allowing the control and verification of the working parameters as shown in the inset (c). Geometrical constraints allow for limited incident angle adjustment within 6.5$^\circ$ from $42^\circ$ side, as shown in inset (d). 

The four major requirements can be summarized as follows:
\begin{enumerate}
\item Electrical control of the UV LED. Electrical signals have to be connected with the auxiliary power supply/monitor using vacuum-grade using custom feed-throughs designed specifically for the mechanical arrangement specific to the instrument.
\item Mechanical Control of the tilt and proximity with the ability to position the instrument within the vacuum chamber ("in use"/"parked" conditions) without fluctuation in vacuum pressure.
\item Technical means of adjusting the instrument relative to the electron beam EB. The end-user solution requires that after initial mechanical boundary setup and installation, the instrument can be freely brought to the proximity of the sample or withdrawn without causing potential mechanical collisions with other equipment or internal parts of the microscope.
\item All materials used must comply with low-outgassing and vacuum-compatible characteristics.
\end{enumerate}

\begin{figure}[ht]
\centering\includegraphics[width=0.9\textwidth]{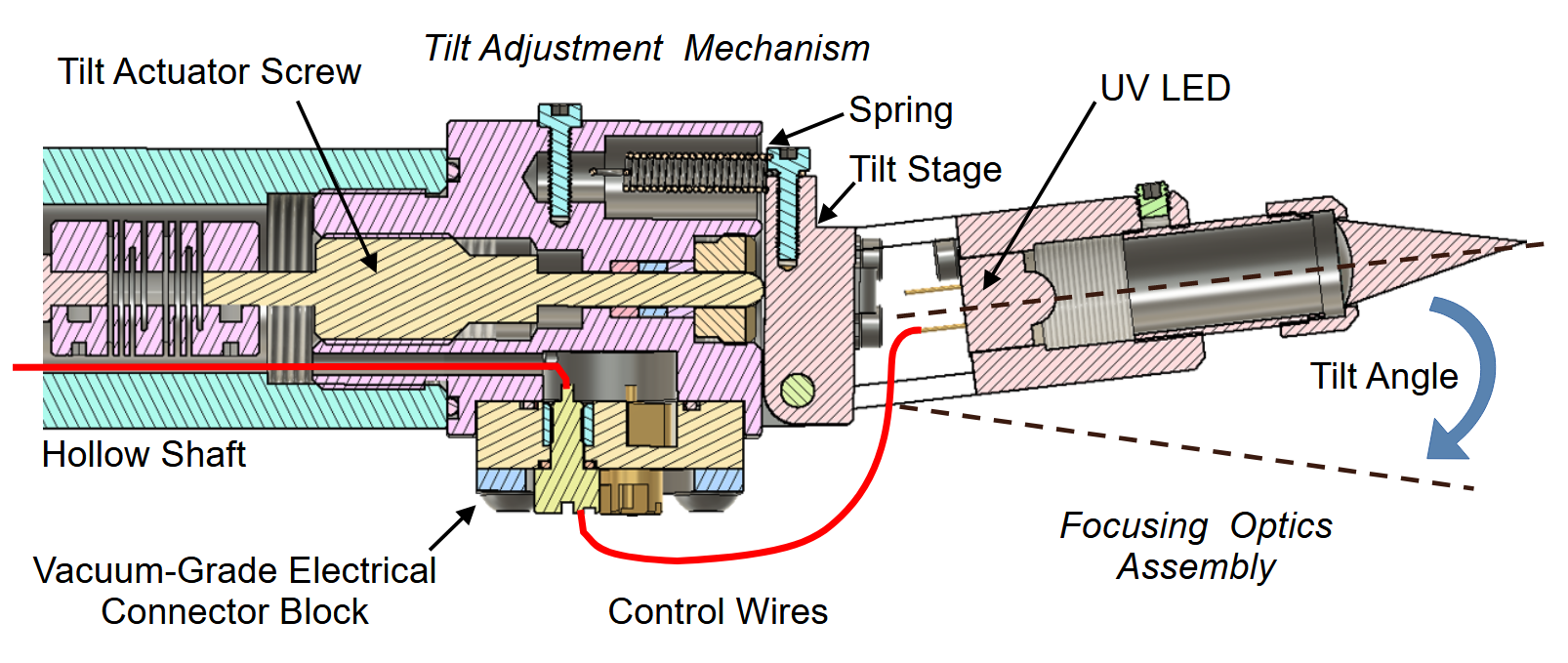}
\caption{\label{f-illum} Cross section of the UV LED illumination end-effector assembly. Focusing lens was added in front of UV-C LED. Some LEDs have lens incorporated into their housing and some not. }
\end{figure}

\begin{figure}[h!]
\centering\includegraphics[width=0.9\textwidth]{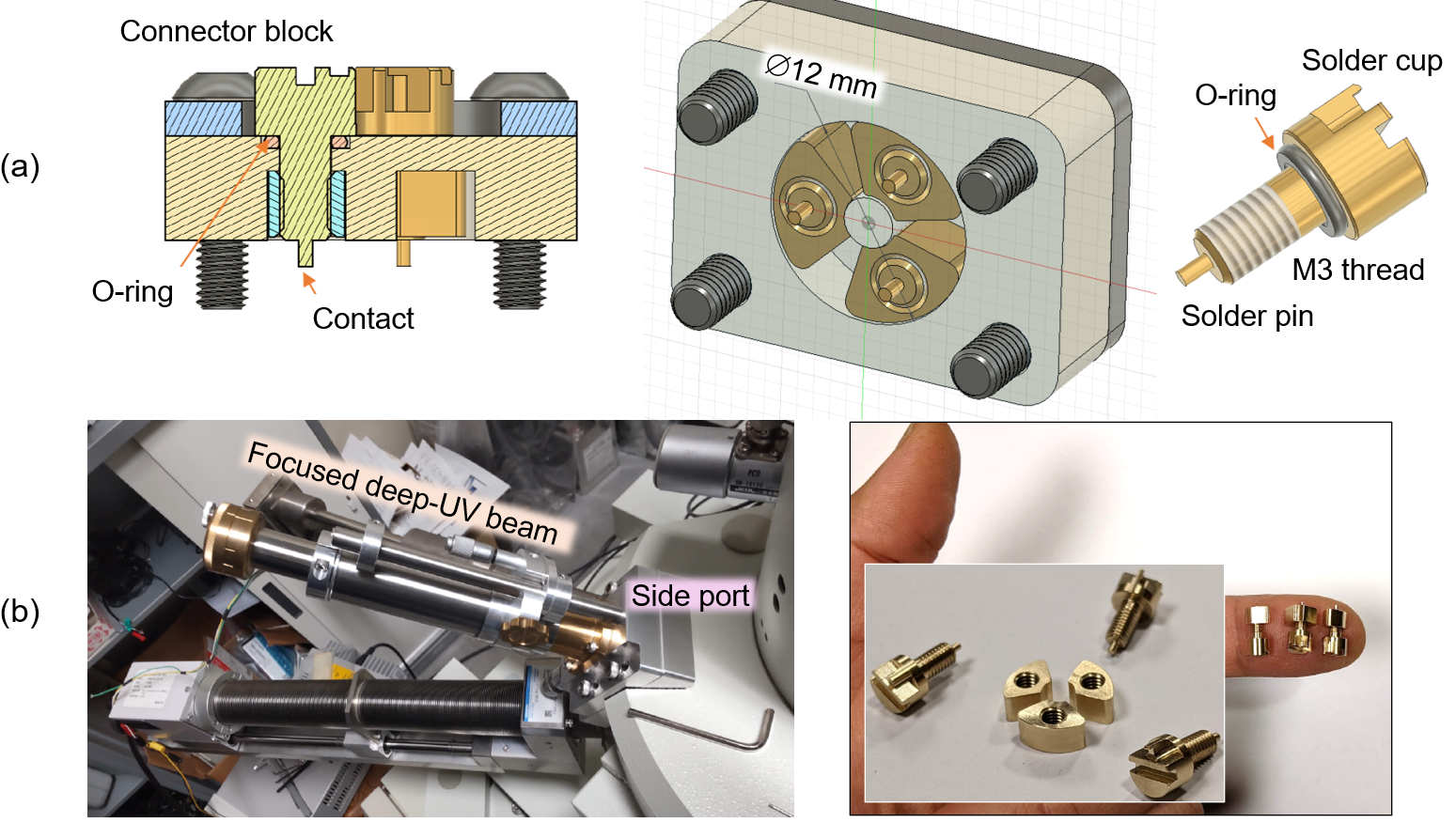}
\caption{\label{f-real} (a) Electrical connector block for a feed-through with 3 electrical contact pins within 12-mm-diameter disk (noteworthy, a standard BNC feed-through has 40-mm-diameter). (b) Actual deep-UV module on a side-port and custom-machined connector parts.}
\end{figure}

To address the above requirements, we have adopted a hollow shaft, telescopic arrangement for the design. This allows for relatively simple vacuum seal and linear bearing arrangement for positioning of the end-effector (the illuminator). The electrical connections as well as the tilt control mechanism are incorporated within the hollow shaft and vacuum-sealed using o-rings incorporated within the mechanism. Moreover, the end-stops of the shaft's travel, its rotational orientation, and lateral alignment can be set and adjusted externally to the vacuum chamber. Figure~\ref{f-illum} shows the cross section of the illuminator. The tilt adjustment mechanism incorporates vacuum seals for the hollow shaft and associated hardware as well as the connector block for the UV source. The tilt stage that regulates the angle of irradiation is adjusted by a screw-operated actuator. The rotating motion of the actuator causes the axial motion of the cylindrical contact point, which causes the angular displacement of the tilt stage. A spring provides a counter-acting force to the actuator providing backlash cancellation.
The LED used in the deep-UV focused beam illuminator was OPTAN 250K-BL (Asahi Kasei Co). The average LED optical power was 3~mW for electrical 10~V and 100~mA. The microlens to focus 15$^\circ$ (approximated) divergent UV-C light onto the sample was Edmund \#48-810 OD9 EF13.5. The focused spot size on the surface of the sample is estimated at 3~mm with calculated irradiance of 38.7 [mW/mm$^2$]. Due to the incident angle, the shape of the spot is elliptical. However, the influence of the elongated shape on the photon-flux variation within SEM-observed area is negligible at high amplification~\cite{Maciek_UV_base}.

Technical innovation was required for combination of miniaturised wire feed-throughs for three electrical connections: two leads to power the LED, and one for the grounding of the instrument. The small size of the end-effector (W30 H35 L80) is crucial to being able to pass through the standard mounting port of the SEM. For a self-contained instrument like the one presented, standard BNC or other electrical feed-throughs are not feasible. Therefore, a custom vacuum-grade miniaturized electrical connector block has been designed and incorporated within the tilt control mechanism. It is specifically designed to allow for a direct current path from the LED connections through the hollow shaft to the external power supply. Each of three contacts is vacuum-sealed by mean's of o-rings (NBR SS-030) to a PTFE body (insulator). The whole contact block assembly is assembled inside the tilt mechanism housing with an o-ring (NBR SS-135) as well, forming a robust miniature solution that is applicable to high vacuum systems and easily scalable.
The fabrication of the components for the prototype instrument was carried out using a combination of machining techniques available in-lab for our group. Rough stock conditioning was performed using a ram-type vertical milling machine (Kanto Koki K2), for round stock processing and bearing/seal interface polishing a manual lathe (Takisawa TSL-500) was used. The fine details on the precision components have been machined using a CNC milling machine developed by our group for lab use using commonly available CNC entry level components. The overall development process has been performed entirely using available low-cost equipment, proving the feasibility, cost-effectiveness, and high potential for academic and industrial applications.

\begin{figure}[h!]
\centering\includegraphics[width=0.9\textwidth]{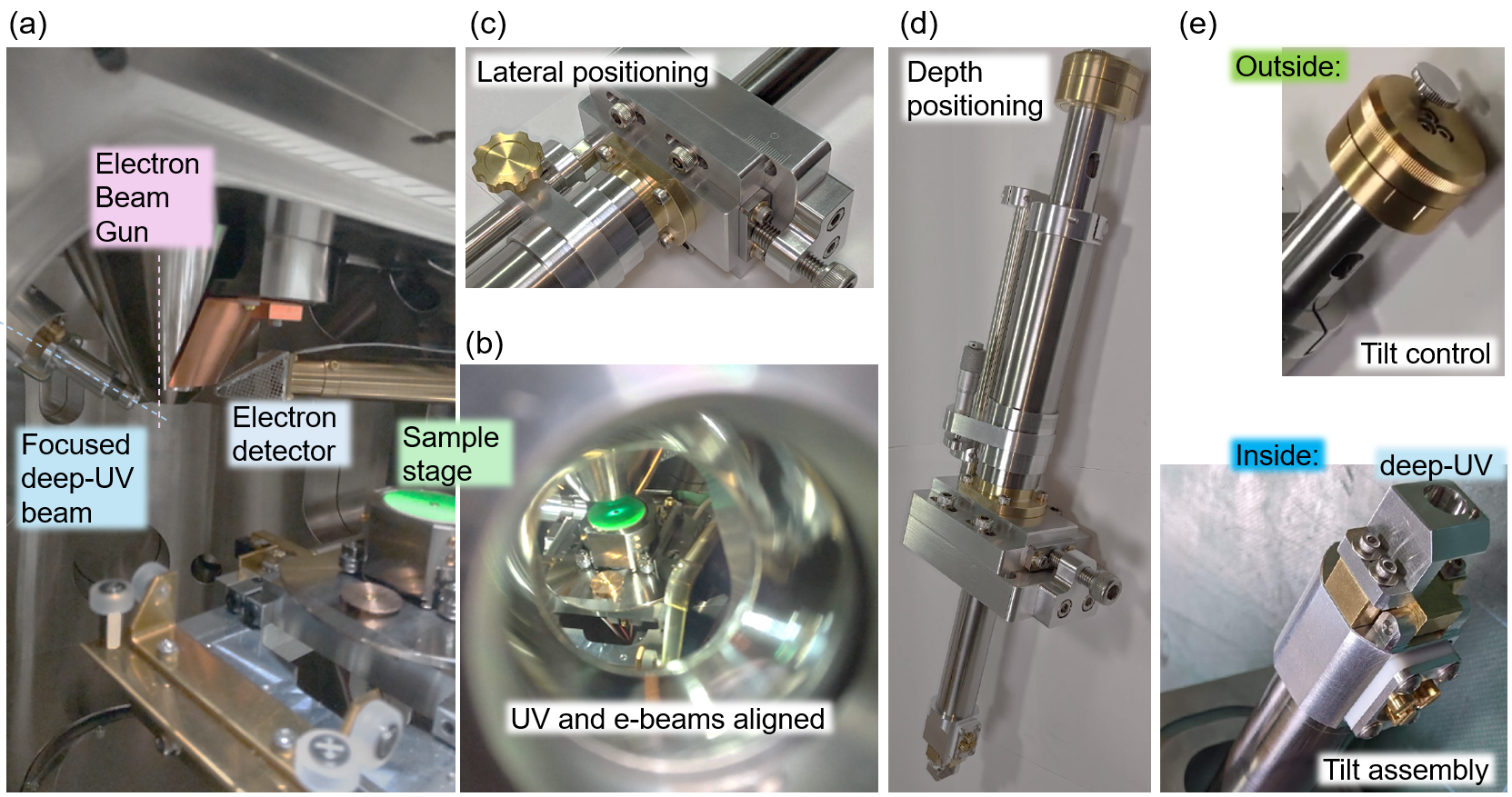}
\caption{\label{f-all} Complete assembly: (a) UV beam and e-beam delivery and detection parts inside SEM chamber, (b) image of aligned UV beam on the centre of the stage (axis of e-beam), (c) lateral UV beam alignment, (d) depth positioning of the UV beam, (e) the tilt control of the UV beam.}
\end{figure}

\section{Results}

\subsection{Realisation of the UV-beam co-illumination}

Figure~\ref{f-real}(a) shows the main connector block which allows electrical control of the internal deep-UV LED. It is custom made and is 3.5 times smaller than the usual electrical BNC feed-through connector. The deep-UV LED can be driven externally by a power supply which modulates DC current with a duty cycle of applied electrical power 10\%(ON)-90\%(OFF). This can provide a proportionally larger UV light intensity on the sample. There are two possible locations on the Jeol 7001 SEM for implementing a practical illumination of the sample during SEM imaging. The one used, shown in Fig.~\ref{f-real}(b), allowed the addition of UV illumination without any disturbance of the electron optics for irradiation of the sample and collection of back-scattered and secondary electrons. 

The instrument assembly of the deep-UV beam control mountable of a side port is shown in Fig.~\ref{f-all}. The modular design of the instrument allows for efficient re-configuration and fine adjustment. The tilt, distance from the sample, and lateral adjustments can be performed while the SEM system is under vacuum without loss of pressure. Auxiliary electrical control of the UV LED allows for current control in the range of 0-100~mA with pre-set duty cycle. The proposed solution is very compact to accommodate the SEM's geometry. The current iteration allows for three electrical connections. However, since the solution is highly scalable, 8-12 connections can be easily implemented by redesigning the connector block module. 

\begin{figure}[hb]
\centering\includegraphics[width=1\textwidth]{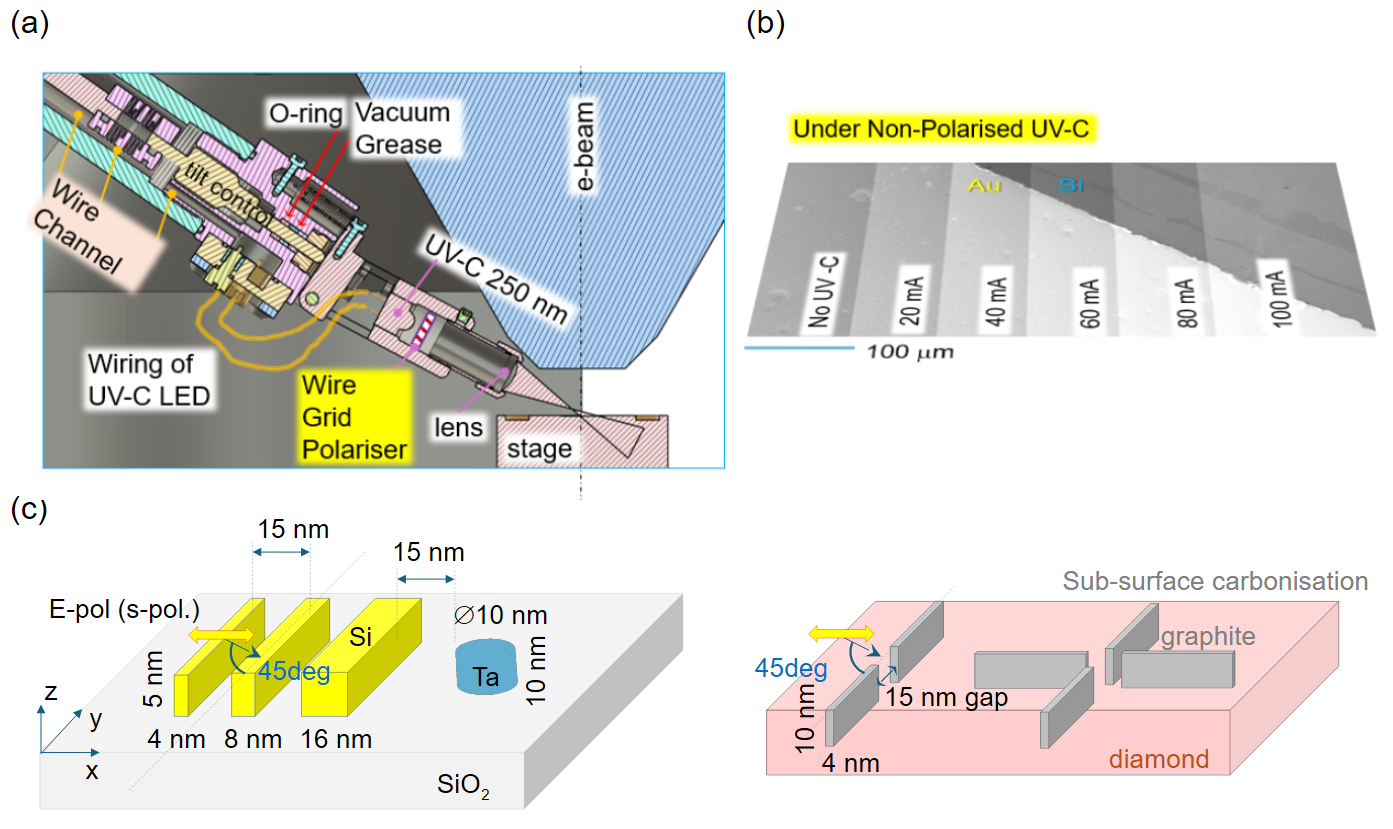}
\caption{\label{f-pol} (a) Cross sectional view of UV-C focused beam with indicated location for 10-mm-diameter wire-grid polariser to produce s-pol. illumination of the sample at $\sim 42^o$. (b) SEM image (slanted view) of the edge between Au ($w_e = 5.1$~eV) coating and Si ($w_e = 4.7$~eV) at different UV-C (4.96~eV) diode currents; light was not polarised.(c) Two models used for numerical simulations showing a typical challenging application to observe SEM: on charging surfaces (without metal coating) and nano-scale surface patterns, e.g., graphite in diamond (\emph{sp$^2$ in sp$^3$} bonding).}
\end{figure}

\subsection{Polarisation control of UV illumination}

With demonstrated feasibility to realise complimentary UV-beam illumination during SEM imaging with intricate control of lateral, axial, and tilt control, the next useful functionality is hypothesized. The implementation of a linear wire-grid polariser in front of the UV LED is a straightforward task (Fig.~\ref{f-pol}(a)). This setup (without the polariser) was used for UV-C enhanced SEM imaging~\cite{Maciek_UV_base} shown in Fig.~\ref{f-pol}(b) where the contrast was tuned with the UV-C LED current (UV intensity on the sample). 

\begin{figure}[hb]
\centering\includegraphics[width=0.8\textwidth]{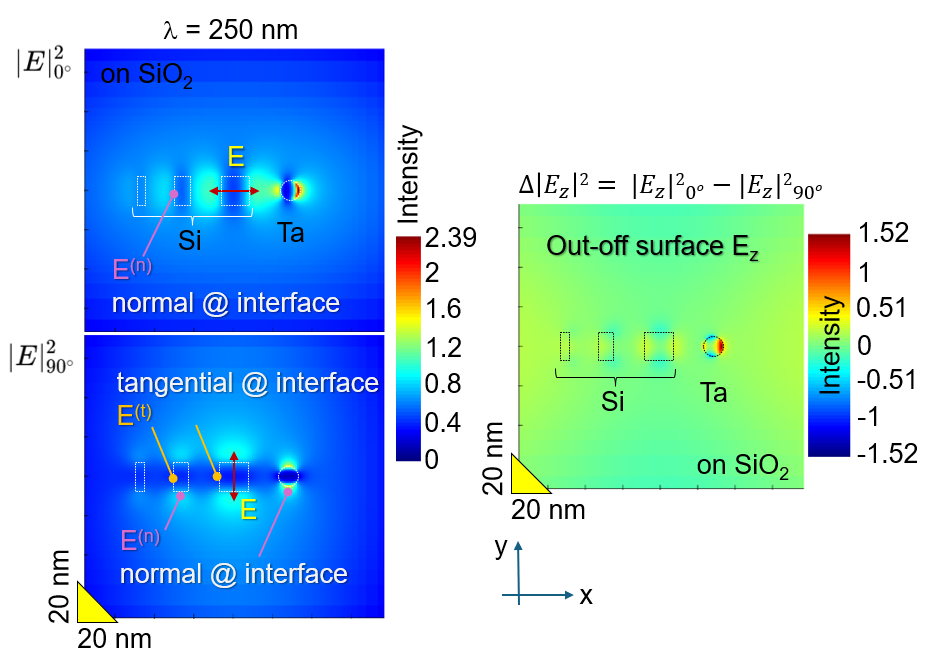}
\caption{\label{f-250} FDTD modelling of total field $|E|^2$ and normal to the surface $|E_z|^2$ components at two orientations of incident polarisation $0^\circ$ and $90^\circ$  at normal incidence as shown in Fig.~\ref{f-pol}(b-top). Boundary conditions are perfectly matching layer (PML). Refractive index at 250 nm for Si (substrate) and Ta (the nano-blocks) are taken from Lumerical, Ansys database. Intensity of the incident light (plane wave) $|E_0|^2=1$. Monitor of the field is 1~nm above the top of the structures (which are 10 nm tall).} 
\end{figure}

First, we consider a simple case when the polarisation of incident UV light is polarised in the sample/stage plane (s-pol. in the plane of incidence). The model pattern of imaging is shown in Fig.~\ref{f-pol}(c), which shows a challenging task for SEM imaging on a dielectric \ce{SiO2} substrate: small feature sized objects with small separation. This is a typical sample which can be encountered in CMOS compatible fabrication.       
There will be UV light-field enhancement at the sharp edges and inside the nano-gaps. This translates to the energy deposition. Since UV photons have energy comparable with that required for the electron escape $h\nu\approx w_e$, the UV enhancement will translate to the emission of secondary electrons. Furthermore, as tunneling ionisation is inherently directional, the relative contribution of the tangential (t) and normal (n) E-field components at the interfaces is crucial. 

Figure~\ref{f-250} shows light intensity distribution in the (x,y)-plane and for the normal to it $|E_z|^2$ component. Finite difference time domain (FDTD, Lumerical, Ansys) was used for the normal incidence to demonstrate specific light intensity enhancement for the normal to the interface E-field component $E^{(n)}$. Depending on the orientation of the sample (e.g., placed on rotational stage), different sides of the Si and Ta nano-blocks are experiencing enhancement of the normal component. Total field enhancement can be doubled due to 250~nm light redistribution of the sub-wavelength pattern with structures close to $\sim 10$~nm feature size. Smaller but considerable E-field enhancement is present for the normal-to-sample's plane component $E_z$, which is not present in the incoming light (paraxial illumination). This $E_z$ field makes a larger portion of the redistributed intensity and can directly contribute to electron emission. In the actual implementation of the UV illumination module at $\sim 45^\circ\pm 7^\circ$ (see Fig.~\ref{f-illum}), the $E_z$ contribution is larger. It would make a methodological sense to record SEM images at four azimuthal angles using tilted UV-illumination. The averaged  SEM image would make representative geometry of the structure and reveal locations for efficient generation of secondary electrons. The FDTD monitor image shown in Fig.~\ref{f-250} is on a plane just 1~nm above the top-of-the-pattern. However, the UV illumination and secondary electrons are emitted from the entire illuminated pattern including the substrate.    

\begin{figure}[h!]
\centering\includegraphics[width=1\textwidth]{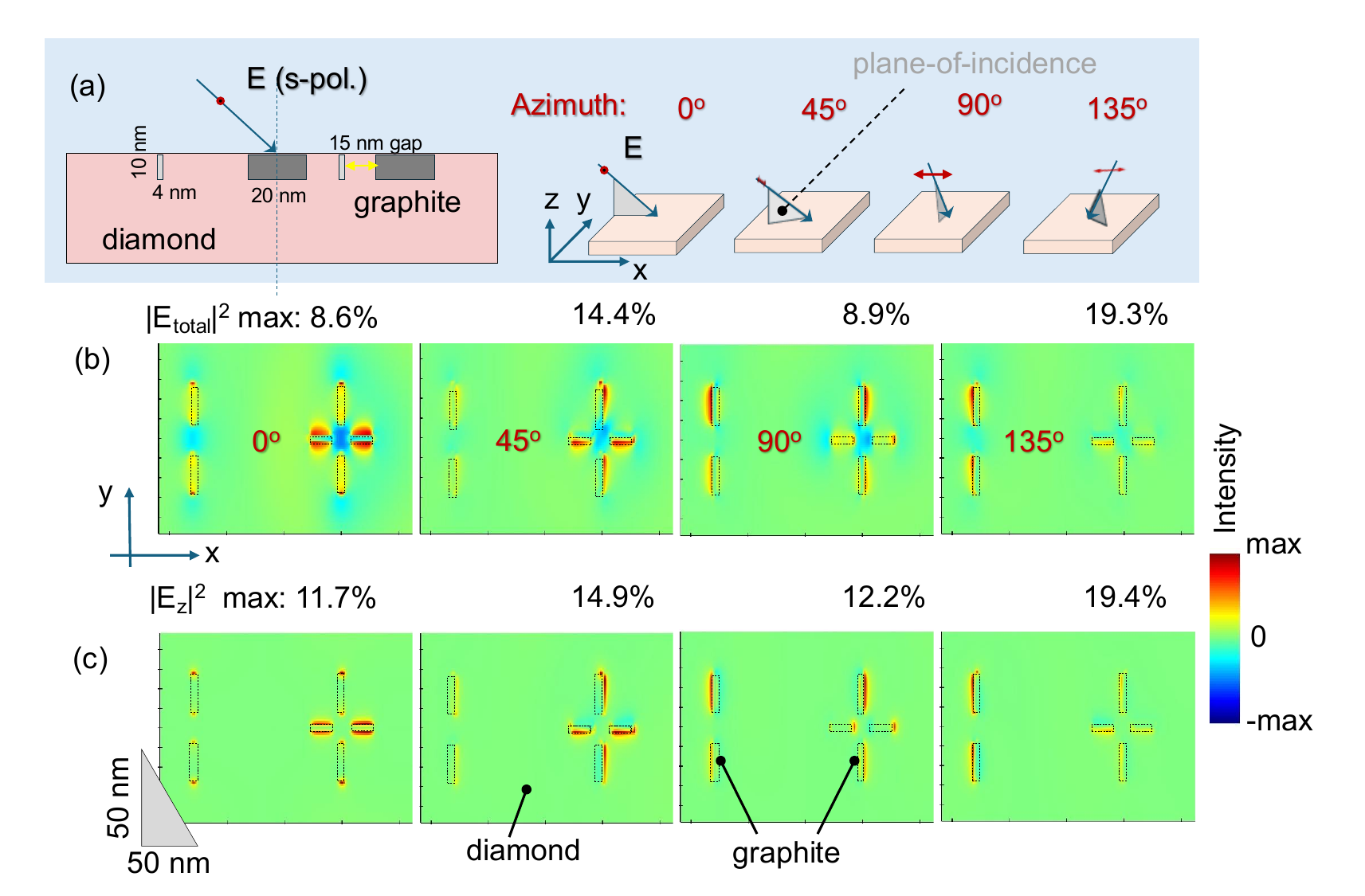}
\caption{\label{f-diam} FDTD modelling of total field $|E|^2$ and normal to the surface $|E_z|^2$ components for four azimuth angles of $0^\circ$, $45^\circ$, $90^\circ$, and $135^\circ$ under s-polarised illumination at the slanted $45^\circ$ incidence onto the pattern as in actual UV illumination module shown in Fig.~\ref{f-pol}(b-bottom); intensity is normalised to the pure diamond substrate by subtraction. The boundary conditions are perfectly matched layers (PML). Refractive index at 250 nm for diamond and graphite are taken from refractiveindex.info database. Intensity of incident light (plane wave) $|E_0|^2=100\%$. Monitor of the field is 1~nm above the surface. The graphite blocks protrude to 10~nm depth below the surface. The outline of graphite rectangular regions are shown by dashed boxes. }
\end{figure}

An example of diamond imaging with nanoscale graphitised planes at slanted incidence of s-pol. 250~nm light at four azimuth angles is shown in Fig.~\ref{f-diam}. The total intensity $|E_{tot}|^2$
in Fig.~\ref{f-diam}(b) and the surface-normal component $|E_{z}|^2$ in Fig.~\ref{f-diam}(c) were normalised to the pure diamond reference by subtraction. The intensity is presented on the same topographic blue-red scale as in Fig.~\ref{f-250}. The positive and negative contrasts in Fig.~\ref{f-diam}(b) indicate redistribution of the incident optical field by the nanoscale graphite structures relative to the pure diamond substrate. In particular, the graphite nanostructures generate a pronounced surface-normal field component $E_z$, as shown in Fig.~\ref{f-diam}(c), which is expected to contribute to secondary electron emission under UV illumination. In the reference pure-diamond case, no structure-induced $E_z$ contrast is present in the monitor plane. Therefore, the subtracted $|E_{z}|^2$ maps in Fig.~\ref{f-diam}(c) directly reflect the contribution from the graphite nanostructures. The modelling shows that a substantial fraction of the redistributed optical field is converted into the surface-normal component $E_z$, particularly near graphite edges and corners. This behaviour is expected to promote secondary electron emission upon UV illumination.

Finally, a rotating control of the mesh-grid polariser would open the possibility to use any orientation of incident polarisation including p-pol. (which is perpendicular to s-pol. discussed so far). Figures~\ref{f-pol-design} and ~\ref{f-p} show the design and modelling, respectively. The diamond surface with nano-blocks of graphite embedded into it was illuminated with p-pol. 250~nm light. Considerably stronger $E_z$-intensity is expected. Therefore, secondary electron emission should be generated. This can be useful for imaging of 2D materials, e.g., graphene, where secondary electrons are excited from the substrate. A single monolayer of graphene ($\sim 0.34$~nm thick) absorbs only 2.3\% of light and is transparent to secondary electrons; the absorption portion $A_0$ is expressed via small structure constant $\alpha$ as $A_0 = \pi\alpha\equiv e^2\pi/\hbar c = \pi/137 \simeq 2.3\%$ where $e$ is the electron charge, $\hbar$ is the small Heisenberg constant, and $c$ is the speed of light.

\section{Discussion}

The proposed addition of linear polarisation of UV-C light polarised in s-pol. (in the plane of incidence) can add new modalities to SEM imagining. Namely, electron photo-emission is strongly directional, which makes enhanced emission for the normal (n) E-field component at the interfaces. The normal component is enhanced at the dielectric (no electron conductivity) interfaces with E-field larger at the side of lower refractive index. The intensity of light $E^2$ is enhanced in proportion of dielectric permittivity squared (hence, refractive index to the power four). Dielectrics with nano-scale features made out of Si, diamond, \ce{Al2O3}, etc., will have the strongest field enhancement. This enhancement facilitates the emission of secondary electrons, which in turn improves the contrast and resolution of the SEM imaging.

One can envisage a possibility of applying the edge-detection method in optical imaging with linearly polarised illumination. The degree of linear polarisation (DOLP) can effectively show the edges of the structures and is calculated from the Stokes vector $(S_0,S_1,S_2,S_3)$ as $DOLP = \sqrt{S_1^2+S_2^2}/S_0$ or from the minimum and maximum intensities as $\frac{I_{max} - I_{min}}{I_{max} + I_{min}}$. The latter expression considers polarisation analysis with a rotating polariser. For linear patterns, it is usually enough to image at two perpendicular polarisations, e.g., $I_{90}$ and $I_{0}$ which will correspond to $I_{max}$ and $I_{min}$. 

Further development of the proposed technique by providing the ability to rotate the angle of linear polarisation \emph{in-situ}, would significantly enhance the imaging capabilities of a standard SEM. Moreover, a higher number of electrical connections would allow for additional instrumentation on the end-effector side. Potentially, several UV irradiation sources (different wavelengths, polarised/un-polarised etc.) could be mounted on the same end-effector concurrently for comparative studies. The presented practical and feasible approach renders a path for further advancements in the field. This would allow for realised p-pol. illumination onto the sample and slanted angle, which promotes directional emission of secondary electrons from the flat surfaces. 

The introduced combined UV-SEM could prove indispensable for characterising new family of materials, the high entropy alloys (HEA), which are crystals with different atoms arranged in a crystalline lattice and have superior mechanical properties~\cite{KIM,SATH,Jung,ZHANG20221}. The open question is "what is their electron work function", $w_e$? For instance, it was found by the density functional theory (DFT) that nano-rough surfaces of Au-Ag alloy result in different binding energies and bond length of nanoalloys~\cite{19jcc925}. This explained efficient surface enhanced Raman scattering (SERS) on Au-Ag and Au-Ag-Cu surfaces~\cite{19jcc925,16sr25010}. Such alloys are used for metamaterials and perfect absorbers in sensor applications~\cite{25e2581} where the reduction of electron emission barrier can be beneficial.   

It could become possible to detect structural defects on nanoscale when they have linear structure/shape~\cite{Maciek_UV_base}, i.e., they are anisotropic in terms of absorption or scattering under UV-C illumination. Such defects are created at the onset of ablation~\cite{26olt115018}. Feasibility to detect anisotropy is based on the discussed normal (n) and tangential (t) boundary conditions for E-field and displacement $D$. The 4+pol. method~\cite{25cbm110573} using linearly polarised light has become a standard procedure to detect anisotropy, moreover, the method can detect presence of anisotropy when spatial resolution is not sufficient at the used wavelength and focusing conditions~\cite{25nse70099}. By harnessing directionality of electron emission via the photo-effect, the virtue of anisotropy detection and analysis by 4+pol. method can be extended in UV-enhanced SEM imaging. The FDTD simulations clearly show edge enhancement effects for UV-illumination for the $E_z$ component.  

\section{Conclusions and Outlook}

The technical implementation of UV-C co-illumination of a sample under SEM imaging is realised with lateral, axial, and tilt control. The electrical wiring is fed through the same side port of the SEM as the instrument module. All moving parts are made with self-lubrication (vacuum grease) and are compatible with SEM vacuum chamber conditions. At the time of writing, the UV co-illumination was installed for 3 years and was functional.   

This study introduces a concept to use the linear polarisation UV-C co-illumination to enhance secondary electrons emission at the interfaces perpendicular (normal E-field component; s-pol.) to the polarisation of the incident UV-C light. Future implementation of the UV-C illumination module in SEM microscopy could have feasible means to rotate the polariser, realising any incident angle between s-pol. and p-pol. The surface of imaged sample, which is normal to the E-field of UV light will produce largest field enhancement at the interface of semiconductor or dielectric. Moreover, an attachment of a UV-illuminator mounted concentrically with the orifice of the e-gun can be envisaged. However, this will release only s-pol. on the sample.

Furthermore, the proposed system offers a versatile platform for nanoscopy of UV-induced structural changes using UV-C light at the wavelength of electron work function as well as adds polarisation controlled contrast. By bridging applications in material science and biology, it enables a deeper fundamental understanding of how light can precisely control material behavior and properties. Nanofilms of 2D materials can be imaged by SEM with co-illumination with UV-C gun for electronic and photonic applications. This is not currently possible for low-conductivity substrates and materials.  

\begin{acknowledgments}
This work was supported by the Japan Science and Technology Agency (JST) through the Core Research for Evolutional Science and Technology (CREST) JPMJCR2003 grant. S.J. is grateful for visiting researcher stay at Shizuoka University and Laser Systems Ltd. and the interest in UV-enhanced electron and ion imaging.
\end{acknowledgments}

\bibliography{aipsamp}

\providecommand{\noopsort}[1]{}\providecommand{\singleletter}[1]{#1}%
\begin{thebibliography}{23}%
\makeatletter
\providecommand \@ifxundefined [1]{%
 \@ifx{#1\undefined}
}%
\providecommand \@ifnum [1]{%
 \ifnum #1\expandafter \@firstoftwo
 \else \expandafter \@secondoftwo
 \fi
}%
\providecommand \@ifx [1]{%
 \ifx #1\expandafter \@firstoftwo
 \else \expandafter \@secondoftwo
 \fi
}%
\providecommand \natexlab [1]{#1}%
\providecommand \enquote  [1]{``#1''}%
\providecommand \bibnamefont  [1]{#1}%
\providecommand \bibfnamefont [1]{#1}%
\providecommand \citenamefont [1]{#1}%
\providecommand \href@noop [0]{\@secondoftwo}%
\providecommand \href [0]{\begingroup \@sanitize@url \@href}%
\providecommand \@href[1]{\@@startlink{#1}\@@href}%
\providecommand \@@href[1]{\endgroup#1\@@endlink}%
\providecommand \@sanitize@url [0]{\catcode `\\12\catcode `\$12\catcode `\&12\catcode `\#12\catcode `\^12\catcode `\_12\catcode `\%12\relax}%
\providecommand \@@startlink[1]{}%
\providecommand \@@endlink[0]{}%
\providecommand \url  [0]{\begingroup\@sanitize@url \@url }%
\providecommand \@url [1]{\endgroup\@href {#1}{\urlprefix }}%
\providecommand \urlprefix  [0]{URL }%
\providecommand \Eprint [0]{\href }%
\providecommand \doibase [0]{https://doi.org/}%
\providecommand \selectlanguage [0]{\@gobble}%
\providecommand \bibinfo  [0]{\@secondoftwo}%
\providecommand \bibfield  [0]{\@secondoftwo}%
\providecommand \translation [1]{[#1]}%
\providecommand \BibitemOpen [0]{}%
\providecommand \bibitemStop [0]{}%
\providecommand \bibitemNoStop [0]{.\EOS\space}%
\providecommand \EOS [0]{\spacefactor3000\relax}%
\providecommand \BibitemShut  [1]{\csname bibitem#1\endcsname}%
\let\auto@bib@innerbib\@empty
\bibitem [{\citenamefont {Gervinskas}, \citenamefont {Seniutinas},\ and\ \citenamefont {Juodkazis}(2013)}]{13lpr1049}%
  \BibitemOpen
  \bibfield  {author} {\bibinfo {author} {\bibfnamefont {G.}~\bibnamefont {Gervinskas}}, \bibinfo {author} {\bibfnamefont {G.}~\bibnamefont {Seniutinas}},\ and\ \bibinfo {author} {\bibfnamefont {S.}~\bibnamefont {Juodkazis}},\ }\bibfield  {title} {\enquote {\bibinfo {title} {Control of surface charge for high-fidelity nanostructuring of materials},}\ }\href {https://doi.org/https://doi.org/10.1002/lpor.201300093} {\bibfield  {journal} {\bibinfo  {journal} {Laser \& Photonics Reviews}\ }\textbf {\bibinfo {volume} {7}},\ \bibinfo {pages} {1049--1053} (\bibinfo {year} {2013})},\ \Eprint {https://arxiv.org/abs/https://onlinelibrary.wiley.com/doi/pdf/10.1002/lpor.201300093} {https://onlinelibrary.wiley.com/doi/pdf/10.1002/lpor.201300093} \BibitemShut {NoStop}%
\bibitem [{\citenamefont {Seniutinas}, \citenamefont {Balčytis},\ and\ \citenamefont {Juodkazis}(2016)}]{16aplp021301}%
  \BibitemOpen
  \bibfield  {author} {\bibinfo {author} {\bibfnamefont {G.}~\bibnamefont {Seniutinas}}, \bibinfo {author} {\bibfnamefont {A.}~\bibnamefont {Balčytis}},\ and\ \bibinfo {author} {\bibfnamefont {S.}~\bibnamefont {Juodkazis}},\ }\bibfield  {title} {\enquote {\bibinfo {title} {Ultraviolet-photoelectric effect for augmented contrast and resolution in electron microscopy},}\ }\href {https://doi.org/10.1063/1.4945357} {\bibfield  {journal} {\bibinfo  {journal} {APL Photonics}\ }\textbf {\bibinfo {volume} {1}},\ \bibinfo {pages} {021301} (\bibinfo {year} {2016})},\ \Eprint {https://arxiv.org/abs/\url{https://pubs.aip.org/aip/app/article-pdf/doi/10.1063/1.4945357/14566956/021301_1_online.pdf}} {\url{https://pubs.aip.org/aip/app/article-pdf/doi/10.1063/1.4945357/14566956/021301_1_online.pdf}} \BibitemShut {NoStop}%
\bibitem [{\citenamefont {Ikhsan}\ \emph {et~al.}(2025)\citenamefont {Ikhsan}, \citenamefont {Masuda}, \citenamefont {Kretkowski}, \citenamefont {Inami},\ and\ \citenamefont {Kawata}}]{Maciek_UV_base}%
  \BibitemOpen
  \bibfield  {author} {\bibinfo {author} {\bibfnamefont {L.~S.}\ \bibnamefont {Ikhsan}}, \bibinfo {author} {\bibfnamefont {Y.}~\bibnamefont {Masuda}}, \bibinfo {author} {\bibfnamefont {M.}~\bibnamefont {Kretkowski}}, \bibinfo {author} {\bibfnamefont {W.}~\bibnamefont {Inami}},\ and\ \bibinfo {author} {\bibfnamefont {Y.}~\bibnamefont {Kawata}},\ }\bibfield  {title} {\enquote {\bibinfo {title} {Contrast enhancement of sem image using photoelectric effect under {UV LED} irradiation},}\ }\href {https://doi.org/10.3390/app152413250} {\bibfield  {journal} {\bibinfo  {journal} {Applied Sciences}\ }\textbf {\bibinfo {volume} {15}} (\bibinfo {year} {2025}),\ 10.3390/app152413250}\BibitemShut {NoStop}%
\bibitem [{\citenamefont {Danilatos}(1980)}]{Danilatos}%
  \BibitemOpen
  \bibfield  {author} {\bibinfo {author} {\bibfnamefont {G.~D.}\ \bibnamefont {Danilatos}},\ }\bibfield  {title} {\enquote {\bibinfo {title} {An atmospheric scanning electron microscope {(ASEM)}},}\ }\href {https://doi.org/https://doi.org/10.1002/sca.4950030314} {\bibfield  {journal} {\bibinfo  {journal} {Scanning}\ }\textbf {\bibinfo {volume} {3}},\ \bibinfo {pages} {215--217} (\bibinfo {year} {1980})},\ \Eprint {https://arxiv.org/abs/https://onlinelibrary.wiley.com/doi/pdf/10.1002/sca.4950030314} {https://onlinelibrary.wiley.com/doi/pdf/10.1002/sca.4950030314} \BibitemShut {NoStop}%
\bibitem [{\citenamefont {Teramoto}\ \emph {et~al.}(2010)\citenamefont {Teramoto}, \citenamefont {Nishiyama}, \citenamefont {Maruyama}, \citenamefont {Konyuuba}, \citenamefont {Abe}, \citenamefont {Guarrera}, \citenamefont {Suga},\ and\ \citenamefont {Sato}}]{Suga}%
  \BibitemOpen
  \bibfield  {author} {\bibinfo {author} {\bibfnamefont {K.}~\bibnamefont {Teramoto}}, \bibinfo {author} {\bibfnamefont {H.}~\bibnamefont {Nishiyama}}, \bibinfo {author} {\bibfnamefont {Y.}~\bibnamefont {Maruyama}}, \bibinfo {author} {\bibfnamefont {Y.}~\bibnamefont {Konyuuba}}, \bibinfo {author} {\bibfnamefont {Y.}~\bibnamefont {Abe}}, \bibinfo {author} {\bibfnamefont {D.}~\bibnamefont {Guarrera}}, \bibinfo {author} {\bibfnamefont {M.}~\bibnamefont {Suga}},\ and\ \bibinfo {author} {\bibfnamefont {C.}~\bibnamefont {Sato}},\ }\bibfield  {title} {\enquote {\bibinfo {title} {Morphological characterization of bacteria using the atmospheric scanning electron microscope (asem)},}\ }\href {https://doi.org/10.1017/S1431927610055625} {\bibfield  {journal} {\bibinfo  {journal} {Microscopy and Microanalysis}\ }\textbf {\bibinfo {volume} {16}},\ \bibinfo {pages} {50--51} (\bibinfo {year} {2010})},\ \Eprint {https://arxiv.org/abs/https://academic.oup.com/mam/article-pdf/16/S2/50/48254561/mam0050.pdf}
  {https://academic.oup.com/mam/article-pdf/16/S2/50/48254561/mam0050.pdf} \BibitemShut {NoStop}%
\bibitem [{\citenamefont {Kleindiek}\ \emph {et~al.}(2016)\citenamefont {Kleindiek}, \citenamefont {Dadras}, \citenamefont {Schock}, \citenamefont {Lieb},\ and\ \citenamefont {Renka}}]{afmsem}%
  \BibitemOpen
  \bibfield  {author} {\bibinfo {author} {\bibfnamefont {S.}~\bibnamefont {Kleindiek}}, \bibinfo {author} {\bibfnamefont {M.}~\bibnamefont {Dadras}}, \bibinfo {author} {\bibfnamefont {K.}~\bibnamefont {Schock}}, \bibinfo {author} {\bibfnamefont {A.}~\bibnamefont {Lieb}},\ and\ \bibinfo {author} {\bibfnamefont {G.}~\bibnamefont {Renka}},\ }\enquote {\bibinfo {title} {Combining {SEM} with {AFM} for in situ correlative microscopy},}\ in\ \href {https://doi.org/https://doi.org/10.1002/9783527808465.EMC2016.5289} {\emph {\bibinfo {booktitle} {European Microscopy Congress 2016: Proceedings}}}\ (\bibinfo  {publisher} {John Wiley \& Sons, Ltd},\ \bibinfo {year} {2016})\ pp.\ \bibinfo {pages} {999--1000},\ \Eprint {https://arxiv.org/abs/https://onlinelibrary.wiley.com/doi/pdf/10.1002/9783527808465.EMC2016.5289} {https://onlinelibrary.wiley.com/doi/pdf/10.1002/9783527808465.EMC2016.5289} \BibitemShut {NoStop}%
\bibitem [{\citenamefont {Kretkowski}\ \emph {et~al.}(2024{\natexlab{a}})\citenamefont {Kretkowski}, \citenamefont {Hosomi}, \citenamefont {Futamata}, \citenamefont {Inami},\ and\ \citenamefont {Kawata}}]{Maciek_EXA_turret}%
  \BibitemOpen
  \bibfield  {author} {\bibinfo {author} {\bibfnamefont {M.}~\bibnamefont {Kretkowski}}, \bibinfo {author} {\bibfnamefont {K.}~\bibnamefont {Hosomi}}, \bibinfo {author} {\bibfnamefont {H.}~\bibnamefont {Futamata}}, \bibinfo {author} {\bibfnamefont {W.}~\bibnamefont {Inami}},\ and\ \bibinfo {author} {\bibfnamefont {Y.}~\bibnamefont {Kawata}},\ }\bibfield  {title} {\enquote {\bibinfo {title} {Fast sample switching mechanism for atmospheric scanning electron microscopy and electron beam irradiation systems of living cells},}\ }\href {https://doi.org/10.12693/APhysPolA.146.382} {\bibfield  {journal} {\bibinfo  {journal} {Acta Physica Polonica A}\ }\textbf {\bibinfo {volume} {146}},\ \bibinfo {pages} {382} (\bibinfo {year} {2024}{\natexlab{a}})}\BibitemShut {NoStop}%
\bibitem [{\citenamefont {Kretkowski}\ \emph {et~al.}(2024{\natexlab{b}})\citenamefont {Kretkowski}, \citenamefont {Katai}, \citenamefont {Futamata}, \citenamefont {Inami},\ and\ \citenamefont {Kawata}}]{MK_Vacuum_Gripper}%
  \BibitemOpen
  \bibfield  {author} {\bibinfo {author} {\bibfnamefont {M.}~\bibnamefont {Kretkowski}}, \bibinfo {author} {\bibfnamefont {J.}~\bibnamefont {Katai}}, \bibinfo {author} {\bibfnamefont {H.}~\bibnamefont {Futamata}}, \bibinfo {author} {\bibfnamefont {W.}~\bibnamefont {Inami}},\ and\ \bibinfo {author} {\bibfnamefont {Y.}~\bibnamefont {Kawata}},\ }\bibfield  {title} {\enquote {\bibinfo {title} {Vacuum gripper-based practical method of gentle deposition of living cells and its filter substrates onto sin films for electron beam irradiation experiments},}\ }in\ \href@noop {} {\emph {\bibinfo {booktitle} {Recent Advances in Technology Research and Education}}},\ \bibinfo {editor} {edited by\ \bibinfo {editor} {\bibfnamefont {Y.}~\bibnamefont {Ono}}\ and\ \bibinfo {editor} {\bibfnamefont {J.}~\bibnamefont {Kondoh}}}\ (\bibinfo  {publisher} {Springer Nature Switzerland},\ \bibinfo {address} {Cham},\ \bibinfo {year} {2024})\ pp.\ \bibinfo {pages} {185--193}\BibitemShut {NoStop}%
\bibitem [{\citenamefont {Guaita}, \citenamefont {Watters},\ and\ \citenamefont {Loerch}(2022)}]{GUAITA2022102484}%
  \BibitemOpen
  \bibfield  {author} {\bibinfo {author} {\bibfnamefont {M.}~\bibnamefont {Guaita}}, \bibinfo {author} {\bibfnamefont {S.~C.}\ \bibnamefont {Watters}},\ and\ \bibinfo {author} {\bibfnamefont {S.}~\bibnamefont {Loerch}},\ }\bibfield  {title} {\enquote {\bibinfo {title} {Recent advances and current trends in cryo-electron microscopy},}\ }\href {https://doi.org/https://doi.org/10.1016/j.sbi.2022.102484} {\bibfield  {journal} {\bibinfo  {journal} {Current Opinion in Structural Biology}\ }\textbf {\bibinfo {volume} {77}},\ \bibinfo {pages} {102484} (\bibinfo {year} {2022})}\BibitemShut {NoStop}%
\bibitem [{\citenamefont {Liu}\ \emph {et~al.}(2026)\citenamefont {Liu}, \citenamefont {Hu}, \citenamefont {Ruterana}, \citenamefont {Olivier}, \citenamefont {Gu},\ and\ \citenamefont {Wang}}]{LIU2026906}%
  \BibitemOpen
  \bibfield  {author} {\bibinfo {author} {\bibfnamefont {B.}~\bibnamefont {Liu}}, \bibinfo {author} {\bibfnamefont {Z.}~\bibnamefont {Hu}}, \bibinfo {author} {\bibfnamefont {P.}~\bibnamefont {Ruterana}}, \bibinfo {author} {\bibfnamefont {E.~J.}\ \bibnamefont {Olivier}}, \bibinfo {author} {\bibfnamefont {L.}~\bibnamefont {Gu}},\ and\ \bibinfo {author} {\bibfnamefont {Y.}~\bibnamefont {Wang}},\ }\bibfield  {title} {\enquote {\bibinfo {title} {Recent research progress of 4d-stem from methodology to application in materials science},}\ }\href {https://doi.org/https://doi.org/10.1016/j.mattod.2025.12.029} {\bibfield  {journal} {\bibinfo  {journal} {Materials Today}\ }\textbf {\bibinfo {volume} {92}},\ \bibinfo {pages} {906--924} (\bibinfo {year} {2026})}\BibitemShut {NoStop}%
\bibitem [{\citenamefont {Troyon}\ \emph {et~al.}(1998)\citenamefont {Troyon}, \citenamefont {Pastré}, \citenamefont {{Pierre Jouart}},\ and\ \citenamefont {{Louis Beaudoin}}}]{TROYON199815}%
  \BibitemOpen
  \bibfield  {author} {\bibinfo {author} {\bibfnamefont {M.}~\bibnamefont {Troyon}}, \bibinfo {author} {\bibfnamefont {D.}~\bibnamefont {Pastré}}, \bibinfo {author} {\bibfnamefont {J.}~\bibnamefont {{Pierre Jouart}}},\ and\ \bibinfo {author} {\bibfnamefont {J.}~\bibnamefont {{Louis Beaudoin}}},\ }\bibfield  {title} {\enquote {\bibinfo {title} {Scanning near-field cathodoluminescence microscopy},}\ }\href {https://doi.org/https://doi.org/10.1016/S0304-3991(98)00049-7} {\bibfield  {journal} {\bibinfo  {journal} {Ultramicroscopy}\ }\textbf {\bibinfo {volume} {75}},\ \bibinfo {pages} {15--21} (\bibinfo {year} {1998})}\BibitemShut {NoStop}%
\bibitem [{\citenamefont {Jirak}, \citenamefont {Čudek},\ and\ \citenamefont {Nedela}(2012)}]{Nedela}%
  \BibitemOpen
  \bibfield  {author} {\bibinfo {author} {\bibfnamefont {J.}~\bibnamefont {Jirak}}, \bibinfo {author} {\bibfnamefont {P.}~\bibnamefont {Čudek}},\ and\ \bibinfo {author} {\bibfnamefont {V.}~\bibnamefont {Nedela}},\ }\bibfield  {title} {\enquote {\bibinfo {title} {Scintillation secondary electron detector for esem and sem},}\ }\href {https://doi.org/10.1017/S1431927612008185} {\bibfield  {journal} {\bibinfo  {journal} {Microscopy and Microanalysis}\ }\textbf {\bibinfo {volume} {18}},\ \bibinfo {pages} {1266--1267} (\bibinfo {year} {2012})}\BibitemShut {NoStop}%
\bibitem [{\citenamefont {Raptis}, \citenamefont {Pikasis},\ and\ \citenamefont {Syvridis}(2016)}]{Raptis:16}%
  \BibitemOpen
  \bibfield  {author} {\bibinfo {author} {\bibfnamefont {N.}~\bibnamefont {Raptis}}, \bibinfo {author} {\bibfnamefont {E.}~\bibnamefont {Pikasis}},\ and\ \bibinfo {author} {\bibfnamefont {D.}~\bibnamefont {Syvridis}},\ }\bibfield  {title} {\enquote {\bibinfo {title} {Power losses in diffuse ultraviolet optical communications channels},}\ }\href {https://doi.org/10.1364/OL.41.004421} {\bibfield  {journal} {\bibinfo  {journal} {Opt. Lett.}\ }\textbf {\bibinfo {volume} {41}},\ \bibinfo {pages} {4421--4424} (\bibinfo {year} {2016})}\BibitemShut {NoStop}%
\bibitem [{\citenamefont {Kim}\ \emph {et~al.}(2026)\citenamefont {Kim}, \citenamefont {Ha}, \citenamefont {Gu}, \citenamefont {Heo}, \citenamefont {Kim}, \citenamefont {Jeong},\ and\ \citenamefont {Kim}}]{KIM}%
  \BibitemOpen
  \bibfield  {author} {\bibinfo {author} {\bibfnamefont {J.}~\bibnamefont {Kim}}, \bibinfo {author} {\bibfnamefont {H.}~\bibnamefont {Ha}}, \bibinfo {author} {\bibfnamefont {G.~H.}\ \bibnamefont {Gu}}, \bibinfo {author} {\bibfnamefont {S.-G.}\ \bibnamefont {Heo}}, \bibinfo {author} {\bibfnamefont {Y.}~\bibnamefont {Kim}}, \bibinfo {author} {\bibfnamefont {G.-H.}\ \bibnamefont {Jeong}},\ and\ \bibinfo {author} {\bibfnamefont {H.~S.}\ \bibnamefont {Kim}},\ }\bibfield  {title} {\enquote {\bibinfo {title} {Argon plasma–induced compressive residual stress for simultaneous strength–ductility enhancement in a cocrfemnni high-entropy alloy},}\ }\href {https://doi.org/https://doi.org/10.1016/j.msea.2026.150062} {\bibfield  {journal} {\bibinfo  {journal} {Materials Science and Engineering: A}\ }\textbf {\bibinfo {volume} {959}},\ \bibinfo {pages} {150062} (\bibinfo {year} {2026})}\BibitemShut {NoStop}%
\bibitem [{\citenamefont {Sathiyamoorthi}\ and\ \citenamefont {Kim}(2022)}]{SATH}%
  \BibitemOpen
  \bibfield  {author} {\bibinfo {author} {\bibfnamefont {P.}~\bibnamefont {Sathiyamoorthi}}\ and\ \bibinfo {author} {\bibfnamefont {H.~S.}\ \bibnamefont {Kim}},\ }\bibfield  {title} {\enquote {\bibinfo {title} {High-entropy alloys with heterogeneous microstructure: Processing and mechanical properties},}\ }\href {https://doi.org/https://doi.org/10.1016/j.pmatsci.2020.100709} {\bibfield  {journal} {\bibinfo  {journal} {Progress in Materials Science}\ }\textbf {\bibinfo {volume} {123}},\ \bibinfo {pages} {100709} (\bibinfo {year} {2022})},\ \bibinfo {note} {a Festschrift in Honor of Brian Cantor}\BibitemShut {NoStop}%
\bibitem [{\citenamefont {Jung}\ \emph {et~al.}(2022)\citenamefont {Jung}, \citenamefont {Han}, \citenamefont {Kim}, \citenamefont {Hidayati}, \citenamefont {Rhyee}, \citenamefont {Lee}, \citenamefont {Kang}, \citenamefont {Choi}, \citenamefont {Jeon}, \citenamefont {Suk},\ and\ \citenamefont {Park}}]{Jung}%
  \BibitemOpen
  \bibfield  {author} {\bibinfo {author} {\bibfnamefont {S.-G.}\ \bibnamefont {Jung}}, \bibinfo {author} {\bibfnamefont {Y.}~\bibnamefont {Han}}, \bibinfo {author} {\bibfnamefont {J.~H.}\ \bibnamefont {Kim}}, \bibinfo {author} {\bibfnamefont {R.}~\bibnamefont {Hidayati}}, \bibinfo {author} {\bibfnamefont {J.-S.}\ \bibnamefont {Rhyee}}, \bibinfo {author} {\bibfnamefont {J.~M.}\ \bibnamefont {Lee}}, \bibinfo {author} {\bibfnamefont {W.~N.}\ \bibnamefont {Kang}}, \bibinfo {author} {\bibfnamefont {W.~S.}\ \bibnamefont {Choi}}, \bibinfo {author} {\bibfnamefont {H.-R.}\ \bibnamefont {Jeon}}, \bibinfo {author} {\bibfnamefont {J.}~\bibnamefont {Suk}},\ and\ \bibinfo {author} {\bibfnamefont {T.}~\bibnamefont {Park}},\ }\bibfield  {title} {\enquote {\bibinfo {title} {High critical current density and high-tolerance superconductivity in high-entropy alloy thin films},}\ }\href@noop {} {\bibfield  {journal} {\bibinfo  {journal} {Nature Communications}\ }\textbf {\bibinfo {volume} {13}},\ \bibinfo {pages} {3373} (\bibinfo
  {year} {2022})}\BibitemShut {NoStop}%
\bibitem [{\citenamefont {Zhang}\ \emph {et~al.}(2022)\citenamefont {Zhang}, \citenamefont {Xu}, \citenamefont {Feng}, \citenamefont {Sun}, \citenamefont {Huang}, \citenamefont {Zhao}, \citenamefont {Yao}, \citenamefont {Chen}, \citenamefont {Lu},\ and\ \citenamefont {Luo}}]{ZHANG20221}%
  \BibitemOpen
  \bibfield  {author} {\bibinfo {author} {\bibfnamefont {N.}~\bibnamefont {Zhang}}, \bibinfo {author} {\bibfnamefont {J.}~\bibnamefont {Xu}}, \bibinfo {author} {\bibfnamefont {Z.}~\bibnamefont {Feng}}, \bibinfo {author} {\bibfnamefont {Y.}~\bibnamefont {Sun}}, \bibinfo {author} {\bibfnamefont {J.}~\bibnamefont {Huang}}, \bibinfo {author} {\bibfnamefont {X.}~\bibnamefont {Zhao}}, \bibinfo {author} {\bibfnamefont {X.}~\bibnamefont {Yao}}, \bibinfo {author} {\bibfnamefont {S.}~\bibnamefont {Chen}}, \bibinfo {author} {\bibfnamefont {L.}~\bibnamefont {Lu}},\ and\ \bibinfo {author} {\bibfnamefont {S.}~\bibnamefont {Luo}},\ }\bibfield  {title} {\enquote {\bibinfo {title} {Shock compression and spallation damage of high-entropy alloy al0.1cocrfeni},}\ }\href@noop {} {\bibfield  {journal} {\bibinfo  {journal} {Journal of Materials Science \& Technology}\ }\textbf {\bibinfo {volume} {128}},\ \bibinfo {pages} {1--9} (\bibinfo {year} {2022})}\BibitemShut {NoStop}%
\bibitem [{\citenamefont {Takenaka}\ \emph {et~al.}(2019)\citenamefont {Takenaka}, \citenamefont {Hashimoto}, \citenamefont {Iwasa}, \citenamefont {Taketsugu}, \citenamefont {Seniutinas}, \citenamefont {Balčytis}, \citenamefont {Juodkazis},\ and\ \citenamefont {Nishijima}}]{19jcc925}%
  \BibitemOpen
  \bibfield  {author} {\bibinfo {author} {\bibfnamefont {M.}~\bibnamefont {Takenaka}}, \bibinfo {author} {\bibfnamefont {Y.}~\bibnamefont {Hashimoto}}, \bibinfo {author} {\bibfnamefont {T.}~\bibnamefont {Iwasa}}, \bibinfo {author} {\bibfnamefont {T.}~\bibnamefont {Taketsugu}}, \bibinfo {author} {\bibfnamefont {G.}~\bibnamefont {Seniutinas}}, \bibinfo {author} {\bibfnamefont {A.}~\bibnamefont {Balčytis}}, \bibinfo {author} {\bibfnamefont {S.}~\bibnamefont {Juodkazis}},\ and\ \bibinfo {author} {\bibfnamefont {Y.}~\bibnamefont {Nishijima}},\ }\bibfield  {title} {\enquote {\bibinfo {title} {First principles calculations toward understanding sers of 2,2'-bipyridyl adsorbed on {Au, Ag, and Au–Ag} nanoalloy},}\ }\href {https://doi.org/https://doi.org/10.1002/jcc.25603} {\bibfield  {journal} {\bibinfo  {journal} {Journal of Computational Chemistry}\ }\textbf {\bibinfo {volume} {40}},\ \bibinfo {pages} {925--932} (\bibinfo {year} {2019})},\ \Eprint
  {https://arxiv.org/abs/https://onlinelibrary.wiley.com/doi/pdf/10.1002/jcc.25603} {https://onlinelibrary.wiley.com/doi/pdf/10.1002/jcc.25603} \BibitemShut {NoStop}%
\bibitem [{\citenamefont {Hashimoto}\ \emph {et~al.}(2016)\citenamefont {Hashimoto}, \citenamefont {Seniutinas}, \citenamefont {Balcytis}, \citenamefont {Juodkazis},\ and\ \citenamefont {Nishijima}}]{16sr25010}%
  \BibitemOpen
  \bibfield  {author} {\bibinfo {author} {\bibfnamefont {Y.}~\bibnamefont {Hashimoto}}, \bibinfo {author} {\bibfnamefont {G.}~\bibnamefont {Seniutinas}}, \bibinfo {author} {\bibfnamefont {A.}~\bibnamefont {Balcytis}}, \bibinfo {author} {\bibfnamefont {S.}~\bibnamefont {Juodkazis}},\ and\ \bibinfo {author} {\bibfnamefont {Y.}~\bibnamefont {Nishijima}},\ }\bibfield  {title} {\enquote {\bibinfo {title} {{Au-Ag-Cu} nano-alloys: tailoring of permittivity},}\ }\href@noop {} {\bibfield  {journal} {\bibinfo  {journal} {Scientific Reports}\ }\textbf {\bibinfo {volume} {6}},\ \bibinfo {pages} {25010} (\bibinfo {year} {2016})}\BibitemShut {NoStop}%
\bibitem [{\citenamefont {Nishijima}\ \emph {et~al.}(2025)\citenamefont {Nishijima}, \citenamefont {Sudo}, \citenamefont {Matsuo},\ and\ \citenamefont {Juodkazis}}]{25e2581}%
  \BibitemOpen
  \bibfield  {author} {\bibinfo {author} {\bibfnamefont {Y.}~\bibnamefont {Nishijima}}, \bibinfo {author} {\bibfnamefont {T.}~\bibnamefont {Sudo}}, \bibinfo {author} {\bibfnamefont {Y.}~\bibnamefont {Matsuo}},\ and\ \bibinfo {author} {\bibfnamefont {S.}~\bibnamefont {Juodkazis}},\ }\bibfield  {title} {\enquote {\bibinfo {title} {A noble metal high-entropy alloy for mid-infrared metasurfaces},}\ }\href {https://doi.org/https://doi.org/10.1016/j.eng.2025.01.017} {\bibfield  {journal} {\bibinfo  {journal} {Engineering}\ }\textbf {\bibinfo {volume} {49}},\ \bibinfo {pages} {81--89} (\bibinfo {year} {2025})}\BibitemShut {NoStop}%
\bibitem [{\citenamefont {Grineviciute}\ \emph {et~al.}(2026)\citenamefont {Grineviciute}, \citenamefont {Huang}, \citenamefont {Mu}, \citenamefont {Nikitina}, \citenamefont {Le}, \citenamefont {Katkus}, \citenamefont {Ang},\ and\ \citenamefont {Juodkazis}}]{26olt115018}%
  \BibitemOpen
  \bibfield  {author} {\bibinfo {author} {\bibfnamefont {L.}~\bibnamefont {Grineviciute}}, \bibinfo {author} {\bibfnamefont {H.-H.}\ \bibnamefont {Huang}}, \bibinfo {author} {\bibfnamefont {H.}~\bibnamefont {Mu}}, \bibinfo {author} {\bibfnamefont {J.}~\bibnamefont {Nikitina}}, \bibinfo {author} {\bibfnamefont {N.~H.~A.}\ \bibnamefont {Le}}, \bibinfo {author} {\bibfnamefont {T.}~\bibnamefont {Katkus}}, \bibinfo {author} {\bibfnamefont {A.~S.~M.}\ \bibnamefont {Ang}},\ and\ \bibinfo {author} {\bibfnamefont {S.}~\bibnamefont {Juodkazis}},\ }\bibfield  {title} {\enquote {\bibinfo {title} {Surface texturing and localized oxidation of tantalum nano-film by fs-laser},}\ }\href {https://doi.org/https://doi.org/10.1016/j.optlastec.2026.115018} {\bibfield  {journal} {\bibinfo  {journal} {Optics \& Laser Technology}\ }\textbf {\bibinfo {volume} {199}},\ \bibinfo {pages} {115018} (\bibinfo {year} {2026})}\BibitemShut {NoStop}%
\bibitem [{\citenamefont {Gassner}\ \emph {et~al.}(2025)\citenamefont {Gassner}, \citenamefont {Vongsvivut}, \citenamefont {Ryu}, \citenamefont {Ng}, \citenamefont {Toplak}, \citenamefont {Anand}, \citenamefont {Takkalkar}, \citenamefont {Fac}, \citenamefont {Sims}, \citenamefont {Wood}, \citenamefont {Tobin}, \citenamefont {Juodkazis},\ and\ \citenamefont {Morikawa}}]{25cbm110573}%
  \BibitemOpen
  \bibfield  {author} {\bibinfo {author} {\bibfnamefont {C.}~\bibnamefont {Gassner}}, \bibinfo {author} {\bibfnamefont {J.}~\bibnamefont {Vongsvivut}}, \bibinfo {author} {\bibfnamefont {M.}~\bibnamefont {Ryu}}, \bibinfo {author} {\bibfnamefont {S.~H.}\ \bibnamefont {Ng}}, \bibinfo {author} {\bibfnamefont {M.}~\bibnamefont {Toplak}}, \bibinfo {author} {\bibfnamefont {V.}~\bibnamefont {Anand}}, \bibinfo {author} {\bibfnamefont {P.}~\bibnamefont {Takkalkar}}, \bibinfo {author} {\bibfnamefont {M.~L.}\ \bibnamefont {Fac}}, \bibinfo {author} {\bibfnamefont {N.~A.}\ \bibnamefont {Sims}}, \bibinfo {author} {\bibfnamefont {B.~R.}\ \bibnamefont {Wood}}, \bibinfo {author} {\bibfnamefont {M.~J.}\ \bibnamefont {Tobin}}, \bibinfo {author} {\bibfnamefont {S.}~\bibnamefont {Juodkazis}},\ and\ \bibinfo {author} {\bibfnamefont {J.}~\bibnamefont {Morikawa}},\ }\bibfield  {title} {\enquote {\bibinfo {title} {Bridging spectroscopy and advanced molecular orientation analysis with new 4+ angle polarization toolbox in quasar},}\
  }\href {https://doi.org/https://doi.org/10.1016/j.compbiomed.2025.110573} {\bibfield  {journal} {\bibinfo  {journal} {Computers in Biology and Medicine}\ }\textbf {\bibinfo {volume} {196}},\ \bibinfo {pages} {110573} (\bibinfo {year} {2025})}\BibitemShut {NoStop}%
\bibitem [{\citenamefont {Ryu}\ \emph {et~al.}(2025)\citenamefont {Ryu}, \citenamefont {Huang}, \citenamefont {Vongsvivut}, \citenamefont {Ng}, \citenamefont {Dumbrytė}, \citenamefont {Narbutis}, \citenamefont {Malinauskas}, \citenamefont {Juodkazis},\ and\ \citenamefont {Morikawa}}]{25nse70099}%
  \BibitemOpen
  \bibfield  {author} {\bibinfo {author} {\bibfnamefont {M.}~\bibnamefont {Ryu}}, \bibinfo {author} {\bibfnamefont {H.-H.}\ \bibnamefont {Huang}}, \bibinfo {author} {\bibfnamefont {J.}~\bibnamefont {Vongsvivut}}, \bibinfo {author} {\bibfnamefont {S.~H.}\ \bibnamefont {Ng}}, \bibinfo {author} {\bibfnamefont {I.}~\bibnamefont {Dumbrytė}}, \bibinfo {author} {\bibfnamefont {D.}~\bibnamefont {Narbutis}}, \bibinfo {author} {\bibfnamefont {M.}~\bibnamefont {Malinauskas}}, \bibinfo {author} {\bibfnamefont {S.}~\bibnamefont {Juodkazis}},\ and\ \bibinfo {author} {\bibfnamefont {J.}~\bibnamefont {Morikawa}},\ }\bibfield  {title} {\enquote {\bibinfo {title} {Anisotropy analysis of bamboo and tooth using 4-angle polarization micro-spectroscopy},}\ }\href {https://doi.org/https://doi.org/10.1002/nano.70099} {\bibfield  {journal} {\bibinfo  {journal} {Nano Select}\ }\textbf {\bibinfo {volume} {n/a}},\ \bibinfo {pages} {e70099} (\bibinfo {year} {2025})},\ \Eprint
  {https://arxiv.org/abs/https://onlinelibrary.wiley.com/doi/pdf/10.1002/nano.70099} {https://onlinelibrary.wiley.com/doi/pdf/10.1002/nano.70099} \BibitemShut {NoStop}%
\end{thebibliography}%
\appendix
\setcounter{figure}{0}\setcounter{equation}{0}
\setcounter{section}{0}\setcounter{equation}{0}
\makeatletter 
\renewcommand{\thefigure}{A\arabic{figure}}
\renewcommand{\theequation}{A\arabic{equation}}
\renewcommand{\thesection}{A\arabic{section}}
\section{Next improvements: rotation of the polariser}

\begin{figure}[ht]
\centering\includegraphics[width=.9\textwidth]{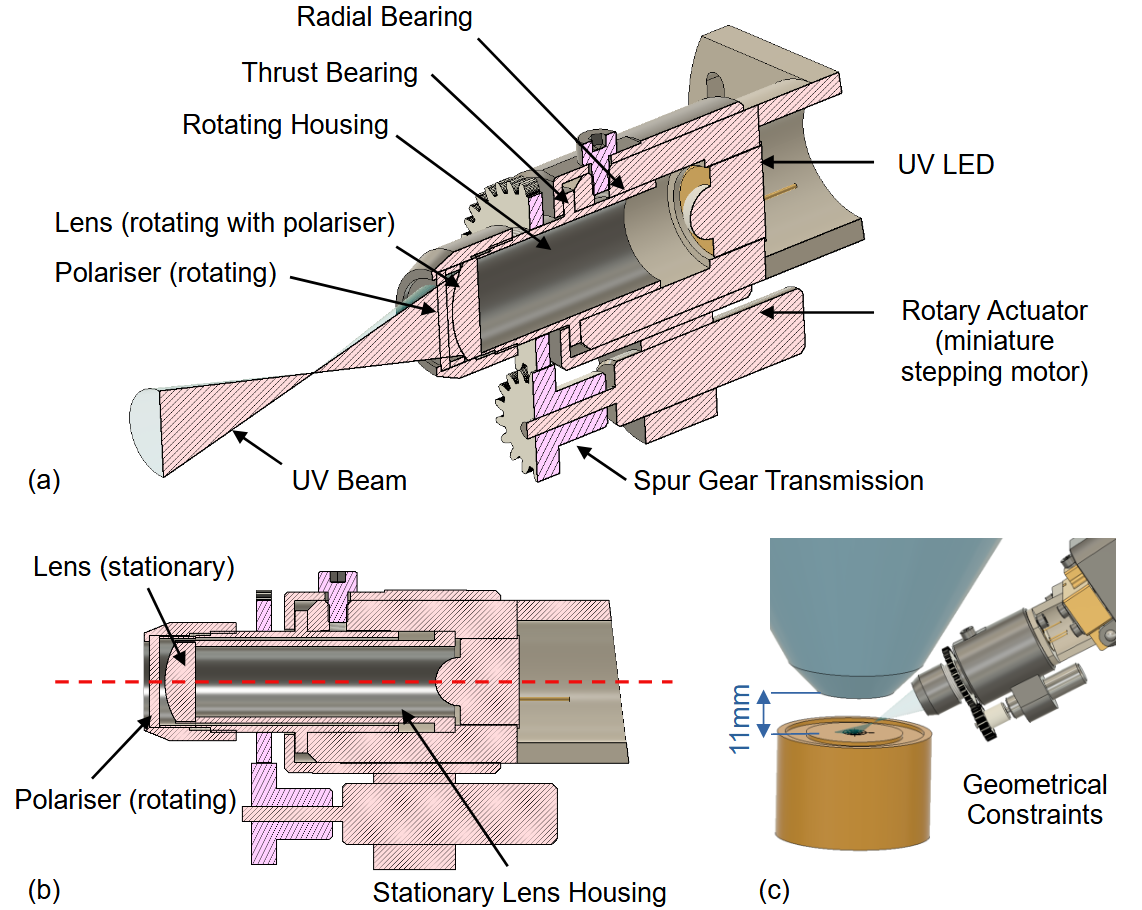}
\caption{\label{f-pol-design} Feasible technical implementations of the polariser rotation. (a) The lens and the polariser are rotated simultaneously. (b) Only the polariser is rotated, while the lens stays stationary. (c) Geometry of the system constraints the physical dimensions of the optics assembly. }
\end{figure}

\begin{figure}[ht]
\centering\includegraphics[width=1\textwidth]{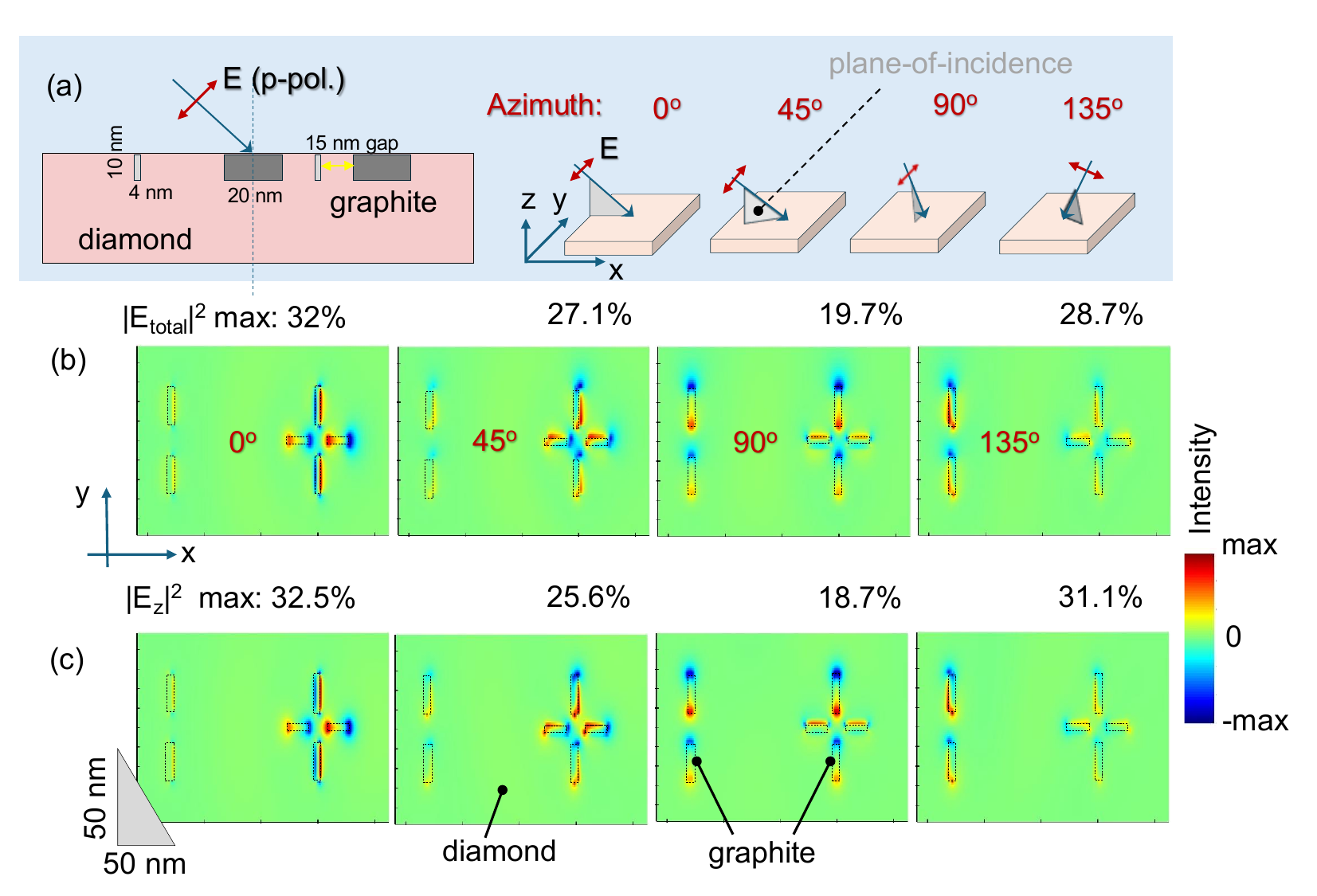}
\caption{\label{f-p} (a) Geometry of model: side-view projection of the sample illuminated by p-pol. UV-C light. FDTD modelling of total field $|E|^2$ (b) and normal to the surface $|E_z|^2$ (c) components at four orientations of incident azimuth angles: $0^\circ$, $45^\circ$, $90^\circ$, and $135^\circ$ under p-polarised illumination at the slanted $45^\circ$ incidence onto the pattern as in actual UV illumination module as shown in Fig.~\ref{f-pol}; intensity is normalised to the pure diamond substrate by subtraction, i.e., the negative value (blue) means that there is less intensity as compared with pure diamond surface. Boundary conditions are perfectly matching layers (PML). Refractive index at 250 nm for diamond and graphite are taken from refractive.info database.Intensity of incident light (plane wave) $|E_0|^2=100\%$. Monitor of the field is 1~nm above the surface. The graphite blocks protrude to 10~nm depth below surface. The outline of graphite rectangular regions are shown by dashed boxes. }
\end{figure}

The practical design of the polarizer rotation mechanism can be achieved in several ways. Due to the geometric constraints of the system and the limited space, two of the most feasible options are proposed (Fig.~\ref{f-pol-design}). In the inset of (a), the proposed mechanism rotates the lens and the polarizer simultaneously. A miniature stepping motor actuator, through a spur gear transmission, rotates the lens/polariser assembly. This solution provides the highest possible optical aperture, ensuring maximum optical power output. However, the light emitted by the LED is not uniform, the projected intensity pattern reflects the emission structure inside the LED. Rotation of the lens may thus be undesirable due to possible fluctuations of the intensity, further emphasized by the imperfections in the optical housing components intrinsic to machining tolerances and processes. The alternative is shown in the inset of (b), where the lens is attached to the housing in a stationary fashion, while only the wire grid polariser is rotated. From an optics perspective, this solution is more appropriate as it mitigates the intensity fluctuations. Taking into consideration the limited space available due to the system restrictions (c), a lens with reduced aperture (by 1.2 mm to 2 mm at least) would have to be used. In this case, the power output will be limited by the smaller aperture and internal reflections in the lens housing. Further research is required to determine the most suitable design with consideration regarding optimal optical power, intensity fluctuations, geometric constraints, cost-performance and maintenance access. In the case of a system, where electron beam physics, optics and mechanics play equal role in driving innovation in SEM imaging, the trade-offs are unavoidable. With the progress of this research, the above factors will be assessed and evaluated for practical implementation purposes.   

With the current state of the UV-C module implementation, all design challenges to have an axially-movable and rotatable stage of the mesh-grid polariser, which is controlled from outside, are technically solved. It is possible to consider not only s-pol. illumination of the sample, but also p-polarised illumination. This directly creates $E_z$-field component, which can be much larger. Figure~\ref{f-p} shows FDTD modeling for p-pol. incidence at the same conventions as in Fig.~\ref{f-diam}.
for s-pol. Larger range of total and $E_z$ intensities is apparent (the blue-red scale); blue region simply means that on surfaces with the buried in graphite nano-blocks the intensity in those regions is lower compared with pure diamond surface. All the high intensity regions have the strongest contribution of the $E_z$ component, which is oscillating perpendicularly to the sample's surface plane and generate secondary electrons. There is orientation dependence with a dipole-like structure in electrically conductive graphite (modeled as $sp^2$ bonded carbon). Similar pattern of light intensity distribution is expected for the 2D materials, e.g., graphene. In 2D material imaging, UV-C light can be useful due to the possibility of generating secondary electrons from the substrate that can still be emitted through the 2D atomic layers. This brings a new modality of UV-enhanced SEM, which is very useful for quantum optics, photonics and electronics of 2D materials.  

The symmetry of the intensity distribution for p-pol. at slanted angle (Fig.~\ref{f-p}) clearly shows the opposite sides of enhanced and decreased intensity compared to the flat diamond surface. This implies that the incidence angle at $0^\circ$ and $180^\circ$ as every other $\pi$-folded pair of angles (e.g., $45^\circ$ and $225^\circ$) has an opposite alternating enhancement pattern. Summing the pair of $\pi$-separated SEM images would correspond to all possible enhancements for secondary electrons contributing to the SEM imaging for the p-pol. UV-C illumination at the particular plane-of-incidence. In this way, the 4-pol. method could be named 8-pol. Imaging of structures at feature sizes in single digits of nanometers in one cross section and elongated in the other, is expected to benefit from the enhancement added by UV-C co-illumination. Further studies and experimental results are planned for the near future.

\end{document}